\documentclass[twoside,twocolumn,9pt]{article}
\usepackage{extsizes}
\usepackage[super,sort&compress,comma]{natbib} 
\usepackage[version=3]{mhchem}
\usepackage[left=1.5cm, right=1.5cm, top=1.785cm, bottom=2.0cm]{geometry}
\usepackage{balance}
\usepackage{mathptmx}
\usepackage{amssymb}   
\usepackage{amsmath}   
\usepackage{sectsty}
\usepackage{graphicx} 
\usepackage{lastpage}
\usepackage[format=plain,justification=justified,singlelinecheck=false,font={stretch=1.125,small,sf},labelfont=bf,labelsep=space]{caption}
\usepackage{float}
\usepackage{fancyhdr}
\usepackage{fnpos}
\usepackage[english]{babel}
\addto{\captionsenglish}{%
  
}
\usepackage{array}
\usepackage{droidsans}
\usepackage{charter}
\usepackage[T1]{fontenc}
\usepackage[usenames,dvipsnames]{xcolor}
\usepackage{setspace}
\usepackage[compact]{titlesec}
\usepackage{hyperref}
\definecolor{cream}{RGB}{222,217,201}

\begin{document}
\pagestyle{fancy}
\thispagestyle{plain}
\fancypagestyle{plain}{
\renewcommand{\headrulewidth}{0pt}
}

\makeFNbottom
\makeatletter
\renewcommand\LARGE{\@setfontsize\LARGE{15pt}{17}}
\renewcommand\Large{\@setfontsize\Large{12pt}{14}}
\renewcommand\large{\@setfontsize\large{10pt}{12}}
\renewcommand\footnotesize{\@setfontsize\footnotesize{7pt}{10}}
\makeatother

\renewcommand{\thefootnote}{\fnsymbol{footnote}}
\renewcommand\footnoterule{\vspace*{1pt}%
\color{cream}\hrule width 3.5in height 0.4pt \color{black}\vspace*{5pt}} 
\setcounter{secnumdepth}{5}

\makeatletter 
\renewcommand\@biblabel[1]{#1}            
\renewcommand\@makefntext[1]%
{\noindent\makebox[0pt][r]{\@thefnmark\,}#1}
\makeatother 
\renewcommand{\figurename}{\small{Fig.}~}
\sectionfont{\sffamily\Large}
\subsectionfont{\normalsize}
\subsubsectionfont{\bf}
\setstretch{1.125} 
\setlength{\skip\footins}{0.8cm}
\setlength{\footnotesep}{0.25cm}
\setlength{\jot}{10pt}
\titlespacing*{\section}{0pt}{4pt}{4pt}
\titlespacing*{\subsection}{0pt}{15pt}{1pt}

\fancyfoot{}
\fancyfoot[RO]{\footnotesize{\sffamily{1--\pageref{LastPage} ~\textbar  \hspace{2pt}\thepage}}}
\fancyfoot[LE]{\footnotesize{\sffamily{\thepage~\textbar\hspace{3.45cm} 1--\pageref{LastPage}}}}
\fancyhead{}
\renewcommand{\headrulewidth}{0pt} 
\renewcommand{\footrulewidth}{0pt}
\setlength{\arrayrulewidth}{1pt}
\setlength{\columnsep}{6.5mm}
\setlength\bibsep{1pt}

\makeatletter 
\newlength{\figrulesep} 
\setlength{\figrulesep}{0.5\textfloatsep} 

\newcommand{\topfigrule}{\vspace*{-1pt}%
\noindent{\color{cream}\rule[-\figrulesep]{\columnwidth}{1.5pt}} }

\newcommand{\botfigrule}{\vspace*{-2pt}%
\noindent{\color{cream}\rule[\figrulesep]{\columnwidth}{1.5pt}} }

\newcommand{\dblfigrule}{\vspace*{-1pt}%
\noindent{\color{cream}\rule[-\figrulesep]{\textwidth}{1.5pt}} }

\makeatother

\twocolumn[
  \begin{@twocolumnfalse}
\vspace{1em}
\sffamily
\begin{tabular}{m{4.5cm} p{13.5cm} }

 & \noindent\LARGE{\textbf{Chiral phase-coexistence in compressed double-twist elastomers}}\\ 
\vspace{0.3cm} & \vspace{0.3cm} \\
& \noindent\large{Matthew P. Leighton$^{a,b}$, Laurent Kreplak$^a$, and Andrew D. Rutenberg$^{a,\ast}$}\\
& \today\\

&  \noindent\normalsize{
We adapt the theory of anisotropic rubber elasticity to model cross-linked double-twist liquid crystal cylinders such as exhibited in biological systems. In mechanical extension we recover strain-straightening, but with an exact expression in the small twist-angle limit. In compression, we observe coexistence between high and low twist phases. Coexistence begins at small compressive strains and is robustly observed for any anisotropic cross-links and for general double-twist functions -- but disappears at large twist angles. Within the coexistence region, significant compression of double-twist cylinders is allowed at constant stress. Our results are qualitatively consistent with previous observations of swollen or compressed collagen fibrils, indicating that this phenomenon may be readily accessible experimentally.
}\\
\end{tabular}
 \end{@twocolumnfalse} \vspace{0.6cm}
  ]

\renewcommand*\rmdefault{bch}\normalfont\upshape
\rmfamily
\section*{}
\vspace{-1cm}

\newcommand{\n}{\hat{\boldsymbol{n}}}
\newcommand{\no}{{\boldsymbol{\hat{n}_0}}}
\newcommand{\strain}{\boldsymbol{\underline{\underline{\lambda}}}}
\newcommand{\xlink}{\boldsymbol{\underline{\underline{\ell}}}}
\newcommand{\lo}{\boldsymbol{\underline{\underline{\ell}}}_0}
\newcommand{\infstrain}{\boldsymbol{\underline{\underline{\epsilon}}}}
\newcommand{\stress}{\boldsymbol{\underline{\underline{\sigma}}}}
\footnotetext{$^a$~Department of Physics and Atmospheric Science, Dalhousie University, Halifax, Nova Scotia, B3H 4R2, Canada.}
\footnotetext{$^b$~Department of Physics, Simon Fraser University, Burnaby, British Columbia, V5A 1S6, Canada.}
\footnotetext{\textit{$^{\ast}$~Corresponding author, email: andrew.rutenberg@dal.ca }}

\section{Introduction}
Chiral nematic (cholesteric) liquid crystals can exhibit a double-twist structure within a cylindrical geometry, in which a molecular director field $\n = -\sin\psi(r) \boldsymbol{\hat{\phi}} + \cos\psi(r) \boldsymbol{\hat{z}}$ has a radius-dependent twist angle $\psi(r)$ with respect to the cylindrical axis $\hat{z}$. Double-twist structures are observed in biological systems such as the keratin macrofibrils in hair or wool \cite{Bryson:2009, Harland:2011}, or the collagen fibrils found in skin, bone, tendon, and the cornea of the eye \cite{Sherman:2015, Raspanti:2018}. They are also found within the ``blue phases'' of chiral liquid crystal systems \cite{Wright:1989}. 

Biological tissues often have substantial amounts of intermolecular cross-links. Enzymatic cross-linking can mechanically \cite{Eekhoff:2018} and thermodynamically \cite{Leighton:2021a} stabilize double-twist collagen fibrils -- and is crucial for healthy tissue formation. Substantial disulfide cross-linking is seen in hair \cite{Matoltsy:1976}.  Non-enzymatic cross-linking, due to advanced glycation endproducts (AGE) \cite{Gautieri:2017}, can also accumulate in various tissues.

The elastomeric theory developed by Warner \textit{et al} \cite{Warner:1996} enables the calculation of mechanical properties of anisotropically cross-linked nematic liquid crystals. Previous work has concentrated on bulk cholesteric liquid crystals, modelling longitudinal strains applied perpendicular to the initial molecular director field -- parallel to the cholesteric twist axis. These systems exhibit a discontinuous director-field reorientation under extension \cite{Warner:2000}, which can indicate a phase transition. Subsequent treatments of cholesteric systems have considered mechanical response to axial strains and electromagnetic fields within the limit of linear elasticity theory \cite{Menzel:2007}, as well as phase transition behaviour under extension and compression with variable chiral solvents \cite{Stille:2009}. Equilibrium phase transitions of cylindrical double-twist elastomers have been considered, \cite{Xing:2008} but without consideration of mechanical strain effects. Other applications have included tunable optical \cite{Finkelmann:2001} or acoustic \cite{Biscari:2014} properties of these systems.

The mechanical Euler buckling of elastic rods on compression is well understood, and can be manipulated by micropatterned materials.\cite{Singamaneni:2010} Phase-coexistence has also been reported under compression for nanostructured materials \cite{Rey:2016} and under extension for macrostructured Kirigami materials \cite{Rafsanjani:2019}. Chiral-shape instabilities of axially compressed elastic rods can also be observed and are well understood.\cite{Miyazaki:1997} However, less attention has been paid to the structure within compressed elastic materials.

Recent experimental work has demonstrated plastic torsional buckling in compressed cross-linked (ex-vivo) collagen fibrils.\cite{Peacock:2020} Since compression in elastomeric systems and the structural changes within compressed elastomeric double-twist cylinders are relatively unexplored, we have explored whether elastomeric theory could help us understand the coupling between chiral structure and mechanical strain in these systems. We find that it does, and that axially compressed double-twist elastomers exhibit a novel chiral phase coexistence. 

Our approach is general. We first compute an expression for the free energy density for a double-twist elastomer, which constitutes a full thermodynamic fundamental relation for the system. Using this fundamental relation as a model we explore the mechanics of double-twist cylinders under both extension and compression. We focus in particular on the changes in molecular orientation as well as the internal stresses within the elastomer. Under sufficient compression we find that large twist angles are always observed -- so we develop a general approach valid for all twist-angle functions $\psi(r)$. We also develop a small-angle approximation that allows us to derive analytic expressions for many properties of collagen fibrils under both extension and compression.  

We limit our numerical studies to two model twist functions -- a rope-like constant twist $\psi=\psi_0$ or a linear twist $\psi = a r$. A constant twist angle has been proposed for corneal collagen fibrils \cite{Ottani:2001, Holmes:2001}. A linear twist structure has been proposed for the cores of blue phases \cite{Wright:1989}, and has been directly observed in hair or wool macrofibrils \cite{Bryson:2009, Harland:2011}. Double-twist behavior that is close to either linear or constant twist is also seen in both equilibrium \cite{Cameron:2020} and non-equilibrium \cite{Leighton:2021a} models of double-twist collagen fibrils.

\section{Model}
When the cross-link configurational entropy dominates, the free-energy density within a nematic liquid crystal elastomer under strain is \cite{Warner:1996}:
\begin{equation}\label{maineq}
f = \frac{1}{2}\mu\textnormal{ Tr} (\lo \strain^\top \xlink^{-1} \strain).
\end{equation}
The energy scale $\mu= k_BT \rho$ is proportional to both temperature $T$ and the volumetric cross-link density $\rho$, where $k_B$ is Boltzmann's constant. The applied deformation gradient tensor is $\strain$. The tensors $\lo$ and $\xlink$ describe the initial and post-strain structure of the elastomer in terms of the initial and post-strain molecular director fields $\no$ and $\n$:
\begin{subequations}
\begin{equation}
\lo =  \boldsymbol{\underline{\underline{\delta}}} + (\zeta - 1)\boldsymbol{\no}\otimes\boldsymbol{\no},
\end{equation}
\begin{equation}
\xlink = \boldsymbol{\underline{\underline{\delta}}} + (\zeta - 1) \n \otimes \n.
\end{equation}
\end{subequations}
Here $\otimes$ indicates a tensor product, and  $\boldsymbol{\underline{\underline{\delta}}}$ denotes the unit tensor. The anisotropy parameter $\zeta$ is the ratio between the cross-link orientation in the directions parallel to and perpendicular to the molecular director field $\n$.  $\zeta$ is typically taken to be greater than 1 in modelling approaches \cite{Warner:1996,Warner:2000}, which is consistent with experimentally observed values of the cross-link anisotropy in nematic liquid crystals \cite{DAllest:1988,Kundler:1998}. 

\begin{figure}[t!] 
\centering
  \includegraphics[width=8.3cm]{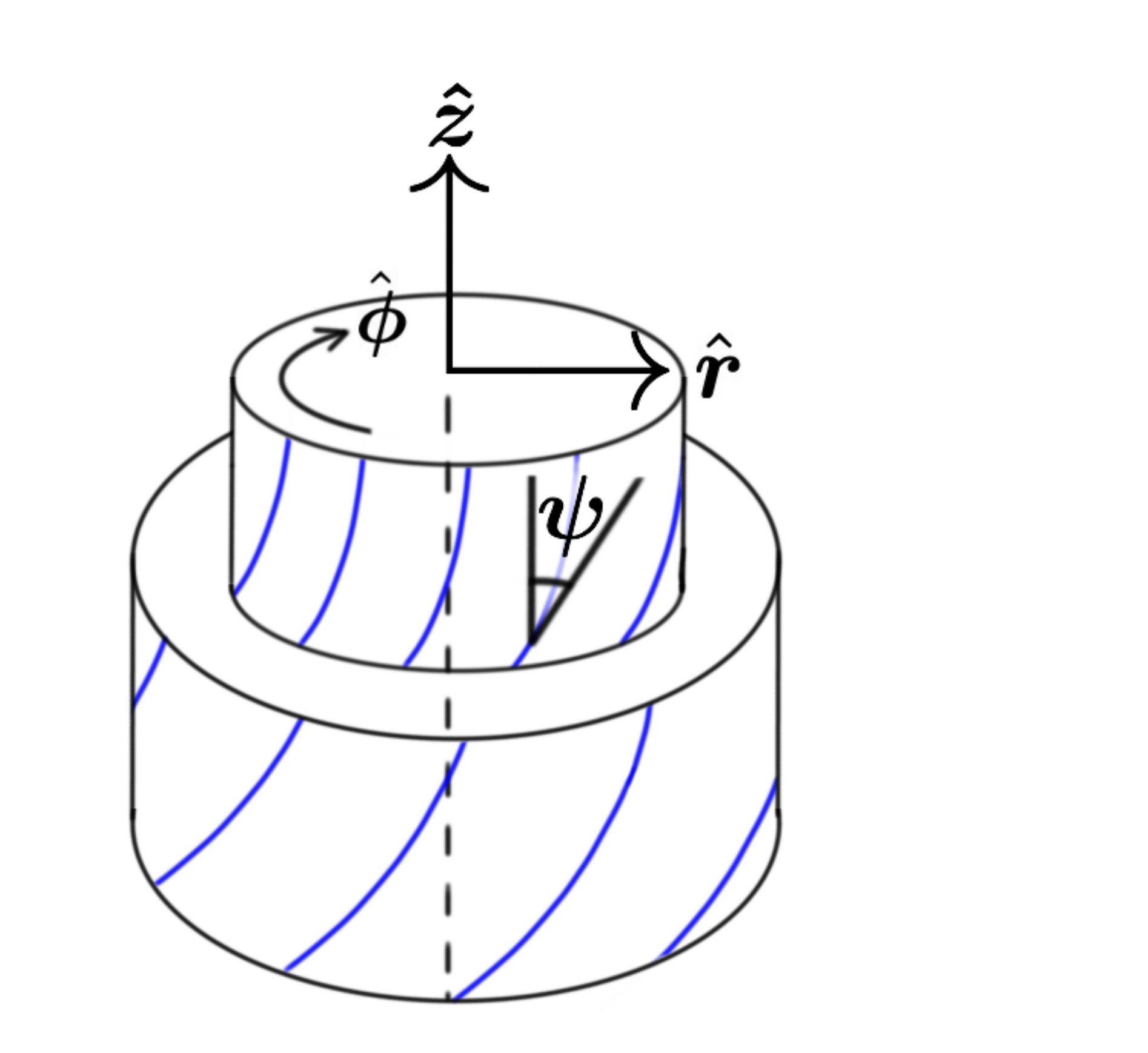}
  \caption{Cutaway view of a double-twist cylinder, with the director field indicated by curved blue lines within the cylinder. The three orthogonal directions in cylindrical coordinates, $\hat{\boldsymbol{r}}$, $\hat{\boldsymbol{\phi}}$, and $\hat{\boldsymbol{z}}$ are indicated. The twist angle $\psi(r)$ depends on the radial distance within the cylinder, but is independent of $\phi$ or $z$. Adapted from \cite{Leighton:2021a}.}
  \label{fig:diagram}
\end{figure}

We assume that both the post-strain and zero-strain director fields retain a double-twist structure, so that $\n = -\sin\psi(r) \boldsymbol{\hat{\phi}} + \cos\psi(r) \boldsymbol{\hat{z}}$ with strain and $\psi_0(r)$ with zero-strain. We consider an extension or compression by a factor of $\lambda$ along the cylinder axis.  We assume that both ends of an incompressible cylinder are fully clamped -- with no shear or rotation. Our deformation is then described by the coordinate transform $z\to \lambda z$, $r\to \lambda^{-1/2}r$, and $\phi\to\phi$. The deformation gradient tensor for this deformation is
\begin{equation}\label{straintensor}
    \strain = \begin{pmatrix} \frac{1}{\sqrt{\lambda}} & 0 & 0\\ 0 & \frac{1}{\sqrt{\lambda}} & 0\\ 0 & 0 & \lambda \end{pmatrix}.
\end{equation}

For small strains, the stress field $\stress$, strain field $\infstrain$, and Helmholtz free energy density $f$ are related by the thermodynamic relation \cite{Landau:1960}    $ \sigma_{ij} = \left( {\partial f}/{\partial \epsilon_{ij}}\right)_{T,N}$,
where the infinitesimal strain tensor is $\infstrain = \frac{1}{2} \left( \strain^\top + \strain\right) - \boldsymbol{\underline{\underline{\delta}}}$.  We are specifically interested in axial (i.e. longitudinal) deformations. The axial stress within the cylinder (or, e.g., fibril) is then given by
\begin{equation} \label{stress}
\sigma = \frac{\partial f}{\partial \lambda},
\end{equation}
where we use a scalar $\sigma$ for simplicity. Similarly we use the scalar $\epsilon \equiv \epsilon_{zz}$, so that the axial strain is $\epsilon = \lambda-1$.

\section{Results}
Using the deformation gradient tensor eqn~\eqref{straintensor}, for general double twist director fields $\no$ and $\n$ we evaluate the free energy density defined in eqn~\eqref{maineq}:
\begin{equation}\label{expanded}
\begin{aligned}
f(r) & = \frac{\mu}{2} \left\{\frac{1}{\lambda}  + \frac{1}{\lambda} [1 + (\zeta-1) \sin^2\psi_0] [1 + (\zeta^{-1}-1) \sin^2\psi] \right.\\
& + \lambda^2 [1 + (\zeta-1) \cos^2\psi_0]  [1 + (\zeta^{-1}-1) \cos^2\psi]\\
&  \left.+ \frac{1}{2} \lambda^{1/2} (2  -\zeta - \zeta^{-1})\sin\big(2\psi_0\big)\sin\big(2\psi\big)\right\}.
\end{aligned}
\end{equation}
The free energy density, $f$, constitutes a thermodynamic fundamental relation for a strained double-twisted liquid-crystal elastomeric cylinder (fibril). The free energy is minimized in equilibrium; this lets us determine the post-strain twist $\psi(r)$ along with the stress within the cylinder.  Our expression for $f$ applies for general double-twist $\psi_0(r)$, extension $\lambda$, and anisotropy $\zeta$. 

The post-strain twist angle function $\psi(r)$ minimizes $f$, so that $\partial f / \partial \psi=0$ for all $r$. This gives
\begin{equation}\label{psieq}
\psi(r) = \frac{1}{2}\cot^{-1}\left( \frac{ (\zeta+1)(\lambda^3-1) + (\zeta-1)(\lambda^3+1)\cos(2\psi_0)}{2\lambda^{3/2}(\zeta-1)\sin(2\psi_0)} \right),
\end{equation}
where we take $\psi\in[0,\pi/2]$. This applies for general $\psi_0(r)$,  $\lambda$, and  $\zeta$.

We can solve equations \ref{stress}, \ref{expanded}, and \ref{psieq} numerically. Common tangent constructions for phase diagrams are performed using custom code.  All of our numerical and plotting code is available on GitHub \cite{github}. While we only show numerical results below for $\zeta>1$, we note that our analytical results above also apply for $\zeta < 1$. 

\subsection{Constant twist coexistence under compression} 
\begin{figure}[t!] 
\centering
  \includegraphics[width=8.3cm]{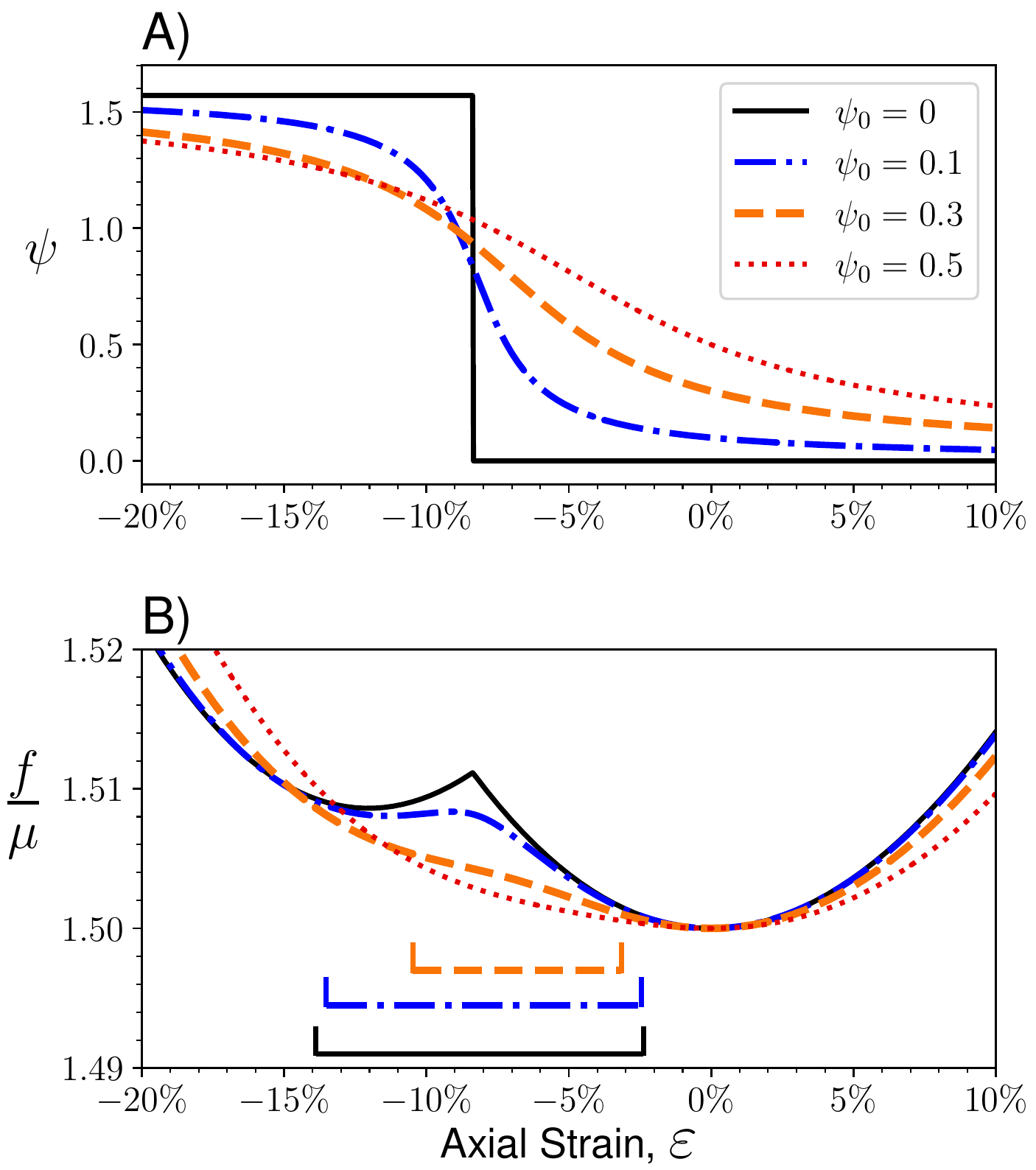}
  \caption{A) The twist angle $\psi$ as a function of longitudinal (axial) strain $\epsilon=\lambda-1$, for various values of $\psi_0$. Positive and negative strains correspond to extension and compression of a cylindrical fibril, respectively. B) The free energy landscape $f/\mu$ as a function of strain for several values of $\psi_0$. Horizontal brackets indicate the coexistence region for corresponding values of $\psi_0$; for $\psi_0=0.5$ there is no coexistence region. In both A) and B) we use a default value of $\zeta=1.3$.}
  \label{fig:energylandscape}
\end{figure}

Using eqn.~\eqref{psieq}, $\psi$ is plotted as a function of strain $\epsilon=\lambda-1$ for various values of constant twist $\psi_0$ in Fig.~\ref{fig:energylandscape}A, with $\zeta=1.3$.  We see that compression monotonically increases $\psi$. 

Fig.~\ref{fig:energylandscape}B shows the corresponding value of the free energy $f$, where we have used the post strain twist angle $\psi$ that minimizes $f$. For sufficiently small $\psi_0$ we find a substantial coexistence region between two compressional strains ($\epsilon_H$ and $\epsilon_L$) that can be identified by a standard common-tangent construction -- with $f'_H = f'_L$ and $f_L =f_H+f'_H (\epsilon_L-\epsilon_H)$. This coexistence allows the system to further reduce the free-energy of the system, and determines thermodynamic equilibrium wherever the free energy is not a convex function of $\lambda$.

\begin{figure}[t!] 
\begin{centering}
  \includegraphics[width=8.3cm]{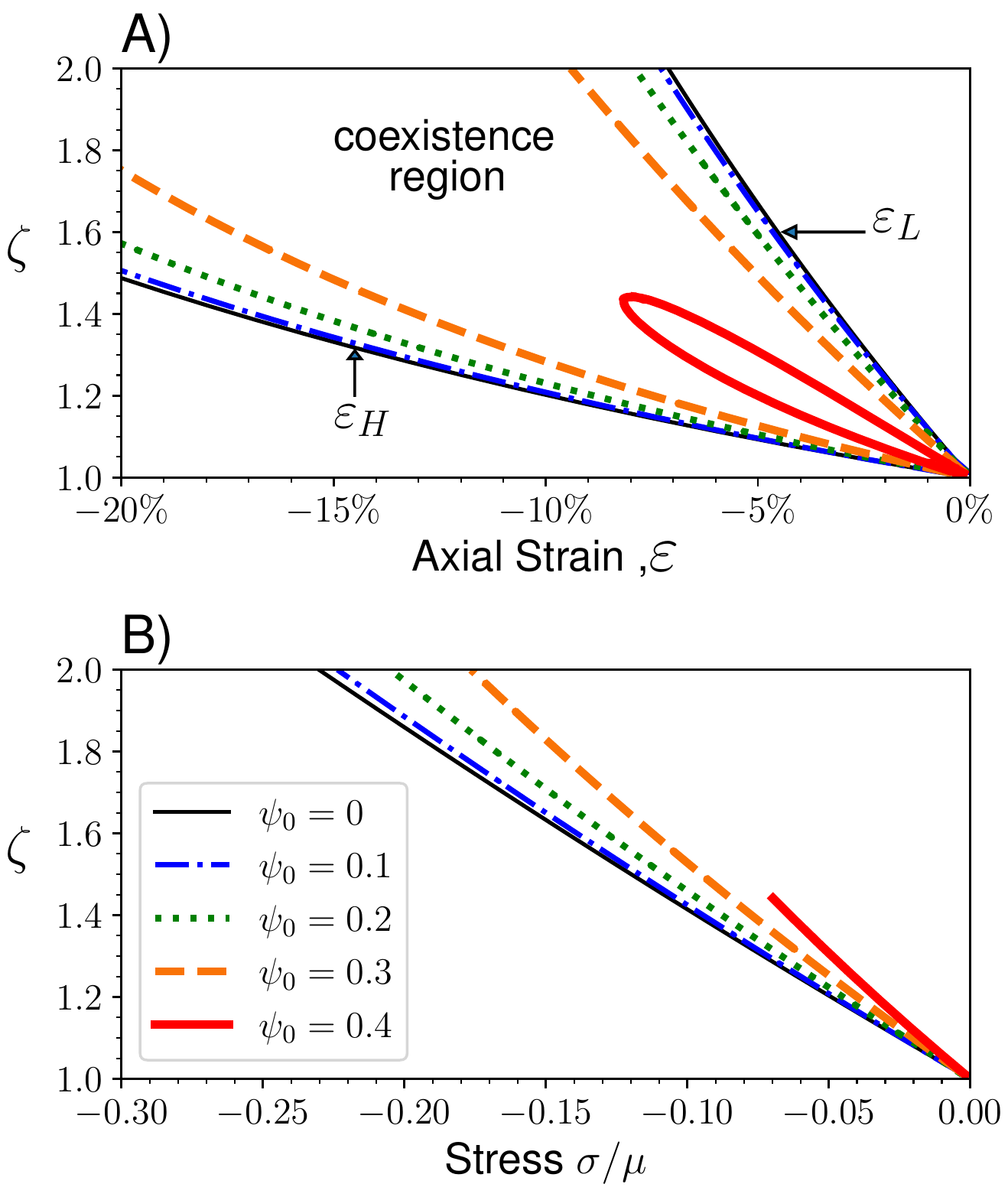}
  \caption{A) Coexistence regions for various values of $\psi_0$, plotted versus cross-linking anisotropy $\zeta$ and strain $\epsilon$. The smaller and larger strain boundary curves of coexistence are indicated  by $\epsilon_L$ and $\epsilon_H$ respectively for $\psi_0=0$. Coexistence is not observed for $\psi_0 > \psi_0^c \approx 0.42$.\cite{multicritical} B) Within the coexistence region stress is constant. The coexistence regions are therefore lines when plotted versus $\zeta$ and stress $\sigma/\mu$.}
  \label{fig:coexistence}
  \end{centering}
\end{figure}

We show the coexistence region in Fig.~\ref{fig:coexistence}A for various values of $\epsilon$, $\psi_0$, and $\zeta$. We observe coexistence at sufficiently small $\zeta$ and $\epsilon<0$ for all $\psi_0 \lesssim 0.42$, while with $\psi_0=0$ we observe coexistence at all $\zeta \neq 1$ and compressive strains $\epsilon < 0$. One consequence of the coexistence across different strains is that the coexistence is also across different twist-angles $\psi$ -- as determined by eqn~\eqref{psieq}.  In Fig.~\ref{fig:coexistence}B we show coexistence curves on the $\zeta-\sigma$ plane for various values of $\psi_0$. As $\psi_0$ increases the extent of the coexistence region decreases -- until it disappears at $\psi_0^c \simeq 0.42385$.\cite{multicritical}

\begin{figure}[t!] 
\centering
   \includegraphics[width=8.3cm]{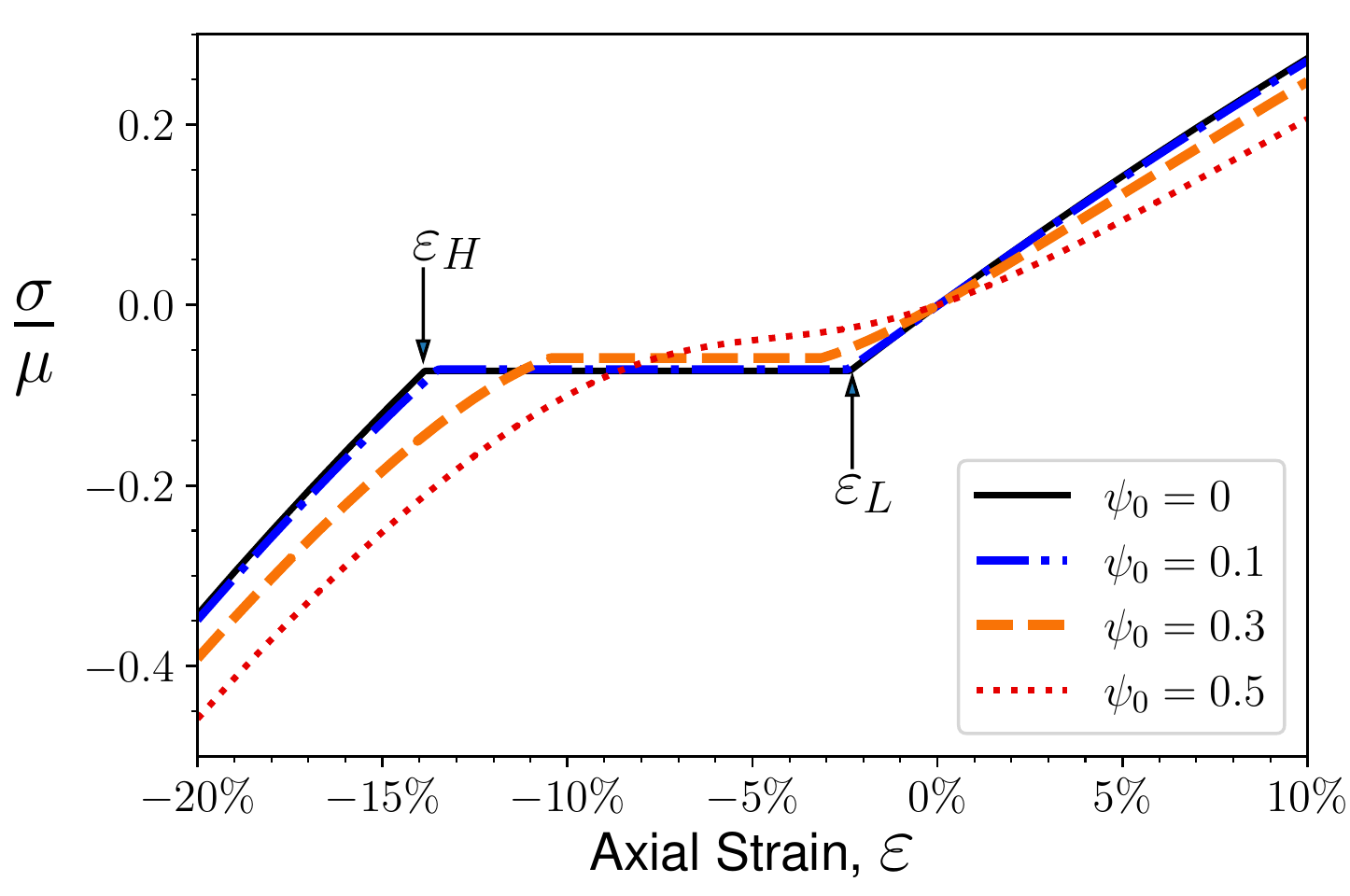}
  \caption{Stress-strain curves of $\sigma/\mu$ vs $\epsilon$ for constant-twist elastomers with various initial twist angles. Coexistence is observed for $\psi_0 \lesssim 0.42$, along with a constant stress between high and low compressive strains: $\epsilon_H$ and $\epsilon_L$, respectively -- indicated with arrows for $\psi_0=0$. Negative strains indicate compression. We use $\zeta=1.3$.}
  \label{fig:pvstate}
\end{figure}

When increasing compressive strains enter the coexistence region, at e.g. $\epsilon_L$ in Fig.~\ref{fig:coexistence}A, a slowly increasing fraction $\chi$ of the system will have local strains $\epsilon_H$ -- while the remainder fraction $1-\chi$ will have unchanging strains at $\epsilon_L$. This is the ``lever rule'' of phase coexistence, and we have that $\chi = (\epsilon-\epsilon_L)/(\epsilon_H-\epsilon_L)$ where $\epsilon$ is the average axial strain. In other words, the free energy $f$ is a linear function of $\epsilon$ within coexistence. This implies that the stress $\sigma = \partial f /\partial \lambda$ is constant within the coexistence region. This constant value was shown in Fig.~\ref{fig:coexistence}B vs $\zeta$. In Fig.~\ref{fig:pvstate} we show the constant coexistence region in a plot of stress $\sigma$ vs strain $\epsilon$. 

\subsection{Linear twist inversion under compression}
To demonstrate that this phase transition behaviour extends to more general double-twist director fields, we also consider a linear twist field $\psi_0(r) = \psi_0^\text{surf} r/R$ -- where $\psi_0^\text{surf}$ is the surface twist and $R$ is the fibril radius. This is a model for, e.g.,  corneal collagen fibrils \cite{Cameron:2020, Leighton:2021a} or keratin macrofibrils \cite{Bryson:2009, Harland:2011}.

\begin{figure}[t!] 
\centering
  \includegraphics[width=8.3cm]{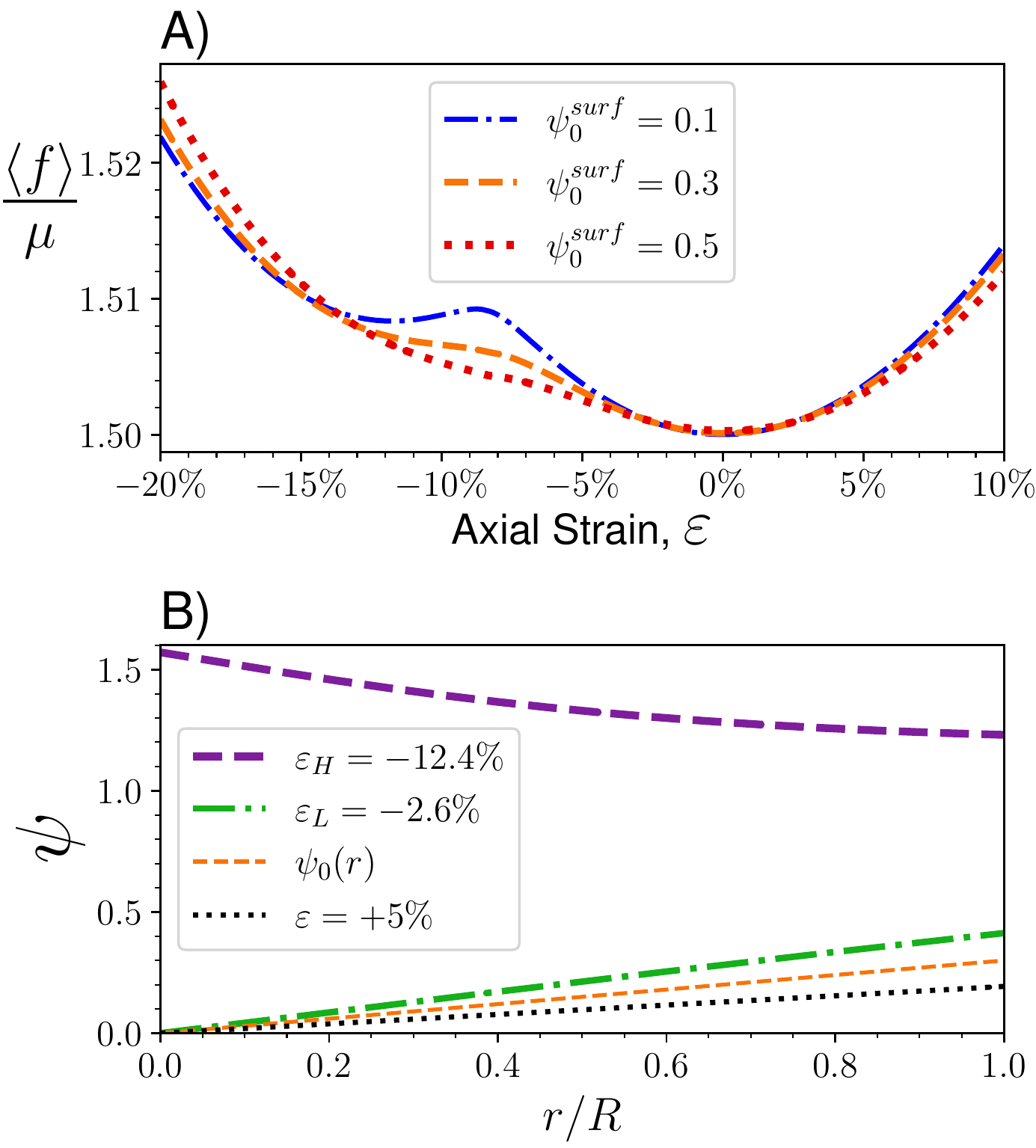} 
  \caption{For cylinders with an initially linear double-twist, $\psi_0(r) = \psi_0^{surf} r/R$. A) The volume-averaged free energy $\langle f \rangle/\mu$ vs longitudinal strain $\epsilon$ for several values of $\psi_0^\text{surf}$ as indicated.   B) The post-strain twist angle function $\psi(r)$  vs $r/R$ for various values of applied strain $\epsilon$, given an initial linear twist function with $\psi_0^\text{surf}=0.3$ (corresponding to the dashed orange curve in A). Note that at $\epsilon_H$ we observe a qualitative inversion of the twist-function. We use $\zeta=1.3$ in both A) and B).}
  \label{fig:lineartwist}
\end{figure}

For this inhomogeneous director field, minimizing the volume averaged free energy density $\langle f \rangle= 2 \int_0^R f(r) rdr /R^2$ still yields eqn~\eqref{psieq}, since $f$ has no dependence on the derivatives of $\psi(r)$.  Fig.~\ref{fig:lineartwist}A shows $\langle f\rangle$ using eqn~\eqref{psieq}. We see that, like in the constant-twist case, the free energy is a non-convex function of the strain for sufficiently small surface twist angles. Thus we still observe phase coexistence in the case of a linear twist function.

Phase-coexistence for linear twist may be unsurprising, given coexistence is observed for similar constant twists. However, near the coexistence region the double-twist function $\psi(r)$ that minimizes $\langle f \rangle$ is no longer linear in $r$ -- as shown by the $\epsilon=\epsilon_H$ curve in Fig.~\ref{fig:lineartwist}B. Furthermore, the twist angle function exhibits striking changes of both monotonicity and convexity under larger compressive strains. This can be clearly understood in the small $\psi_0$ limit.

\subsection{Small $\psi_0$ limit}
The small $\psi_0$ limit is useful for building intuition for the system, and is also observed in e.g. collagen fibrils \cite{Leighton:2021a, Leighton:2021b}. From eqn~\eqref{psieq}, we can extract the leading behavior for small $\psi_0$:
\begin{equation}\label{smallpsi}
\psi(r) = \begin{cases}
    &\ \ \ \psi_0(r) (\zeta-1) \lambda^{3/2}/\left(\zeta\lambda^3 -1 \right),\ \ \ \ \ \text{if $\zeta \lambda^3 >1$,} \\
        \pi/2&-\ \psi_0(r) (\zeta-1) \lambda^{3/2}/\left(\zeta\lambda^3 -1 \right),\ \ \ \ \ \text{if $\zeta \lambda^3 <1$,}
        \end{cases}
\end{equation}
where the corrections are $O(\psi_0^3)$.
This applies to any twist field as long as $\psi_0(r) \ll 1$ for all $r$, which explains the observed twist inversion seen in Fig.~\ref{fig:lineartwist}B. 

When $\psi_0=0$, we can illustrate the phase-coexistence calculation and explicitly show that coexistence is observed for all values of $\zeta \neq 1$. From eqns~\eqref{expanded} and \eqref{smallpsi}, we have
\begin{equation}\label{fcases}
f/\mu = \begin{cases}
     1/\lambda_L+\lambda_L^2/2, & \text{if $\zeta \lambda^3 >1$,} \\
      (1+\zeta^{-1})/(2 \lambda_H) + \lambda_H^2 \zeta/2,& \text{if $\zeta \lambda^3 <1$,}
        \end{cases}
\end{equation}
where we have used $\lambda_L$ for the low-twist small-strain branch and $\lambda_H$ for the high-twist large-strain branch. The common tangent construction for coexistence is given by $f'(\lambda_L) = f'(\lambda_H)$ (which also determines the constant $\sigma=f'$ during coexistence), together with $f(\lambda_L) - \sigma \lambda_L=f(\lambda_H) - \sigma \lambda_H$:
\begin{eqnarray}\label{smallcoexist}
-2/\lambda_L^2 +2 \lambda_L & =& -(1+\zeta^{-1})/\lambda_H^2 + 2 \lambda_H \zeta, \nonumber \\
4/\lambda_L - \lambda_L^2 & =& 2(1+\zeta^{-1})/\lambda_H - \lambda_H^2 \zeta.
\end{eqnarray}
These equations are easily solved numerically, and corresponds to the thin black curves in Fig.~\ref{fig:coexistence}. 

We can take the $\zeta \rightarrow \infty$ limit in eqns~\eqref{smallcoexist} and obtain $\lambda_H = (-6+2 \sqrt{10})^{1/3} \zeta^{-1/3} \simeq 0.69 \zeta^{-1/3}$, $\lambda_L = 2 \lambda_H/(4-\sqrt{10}) \simeq 1.64 \zeta^{-1/3}$, and so constant $\sigma/\mu = -2/\lambda_L^2 \simeq -0.74 \zeta^{2/3}$ within the coexistence region. This illustrates that coexistence can always be observed under compression, even for $\zeta \rightarrow \infty$, for small enough $\psi_0$.

Under extension (and with $\zeta>1$), initial twist-angles always decrease as $\lambda$ increases  -- they exhibit strain-straightening. This can be seen generally in eqn~\eqref{psieq}, and also in the small-angle limit in eqn~\eqref{smallpsi}. We can therefore self-consistently take the small angle limit of $f$ in eqn~\eqref{expanded} and use eqn~\eqref{stress} to obtain the stress vs extensive strain:
\begin{equation}
\begin{aligned}
\sigma & = \mu\left(\lambda - \lambda^{-2}\right) \\
& - \mu \frac{(\zeta-1)(\lambda-1)(\lambda^2+\lambda+1)\left( \lambda^3 [2\zeta (\lambda^3+2) - 5]- 1\right)}{2\lambda^2 (\zeta\lambda^3 - 1)^2}\psi_0^2\\
& + \mathcal{O}(\psi_0^4).
\end{aligned}
\end{equation}
We see that for small initial twist angles under extension, the leading behavior agrees with standard isotropic rubber elasticity. We can also estimate the Young's modulus $E  = d \sigma/d \lambda$, and obtain
\begin{equation} \label{youngs}
E  = 3 \mu (1-3 \psi_0^2) + O(\psi_0^4) 
\end{equation}
at $\lambda=1$.

\section{Discussion and Conclusions}
We have considered the effects of axial strain on double-twist elastomeric cylinders. We have focused on the relatively simple case of a constant twist angle, like a twisted rope, though our results also apply to twist angles that depend on radial distance from the cylinder center. Minimizing the standard entropic free-energy arising from anisotropic cross-linking, we obtain standard strain-straightening under axial extension --  with a simple analytic form at small twist angles.

Under axial compression, we have identified a novel phase-coexistence between high and low twist double-twist phases that begins at small compressive strains. This phase-coexistence is observed for initial twist angles up to $\psi_0 \simeq 0.42$ and for all non-zero values of the cross-link anisotropy parameter $\zeta$. Notably, even an initially achiral cylinder with $\psi_0=0$ will spontaneously exhibit a strong chiral rotation at moderate compressive strains. We would expect physical achiral systems to exhibit spontaneous chiral symmetry breaking as a result. 

The chiral double-twist instabilities we describe here appear to be novel. In particular, they are evident even in an initially achiral cylinder with $\psi_0=0$. Mechanical transitions in double-twist elastomers have not been previously studied, to our knowledge. \citet{Xing:2008} showed that a double-twist configuration was a good candidate ground state for an unstressed initially isotropic elastomeric cylinder that was moved into a cholesteric phase – i.e. due to changing Frank free-energy contributions. We have not included Frank free-energy terms in our treatment. 

Mechanical transitions of strained elastomers have long been studied, but typically under extensional strain.\cite{Mitchell:1993, Warner:1996, Warner:2000} Both axial extension \cite{Burridge:2006, Stille:2009} and compression \cite{Stille:2009} of bulk cholesteric elastomers have been studied. Singular transitions were predicted, though coexistence was not explored. Since these transitions relied on relatively weak Frank free-energy contributions, they may be less robust experimentally than the elastomeric transitions and coexistence we report.

Ex-vivo collagen fibrils appear to be well described by a double-twist configuration, and to be in the strongly cross-linked regime that should be dominated by our elastomeric model \cite{Leighton:2021a}. Recent experimental studies of collagen fibrils under both extension \cite{Bell:2018} and compression \cite{Peacock:2020} appear to qualitatively support our results. Under axial extension, \citet{Bell:2018} observed strain-straightening of the average twist of corneal fibrils. This is consistent with Fig.~\ref{fig:energylandscape} and eqn~\eqref{smallpsi}. A more detailed analysis of the differences between fibril strain and D-band strain in collagen fibrils under extension is in progress \cite{Leighton:2021b}.

Experimentally, compressed ex-vivo (cross-linked) collagen fibrils attached to elastic substrates exhibit buckled regions that coexist with unbuckled regions along the fibril length \cite{Peacock:2020}. Qualitatively consistent with our results is that this coexistence starts  at small compressional strains (below $1\%$). Also consistent with coexistence, the amplitude of the buckled regions does not appear to increase with compressional strain but their frequency of appearance does. Furthermore, the buckled regions exhibit an increased fibril diameter consistent with the smaller $\lambda$ and increased transverse $1/\sqrt{\lambda}$ predicted from our results in Fig.~\ref{fig:coexistence} and eqn~\eqref{straintensor}. 

These AFM studies of compressed collagen fibrils \cite{Peacock:2020} were not able to resolve the fibril twist. However, much earlier EM studies of swollen fibrils due to urea treatment \cite{Lillie:1977} did exhibit the coexistence of strongly twisted swollen regions of the fibril with less twisted narrower regions. While these were not studies of compressed fibrils, the qualitative similarity to our proposed coexistence indicates that a similar phenomenon may be observed as osmotic pressure is varied. 

We can use our coexistence region plot Fig.~\ref{fig:coexistence}A to approximate $\zeta \approx 1.1$ for $\epsilon \approx -0.01$ from \citet{Peacock:2020} -- where these are experimental upper-bounds for collagen fibrils from tendon (with $\psi_0 \simeq 0.1$). This implies from Fig.~\ref{fig:coexistence}B that the compressive stress within coexistence is given by $\sigma \approx 0.02 \mu$. Since \citet{Peacock:2020} reported irreversible plastic damage, we can take this as an upper bound of the yield stress of collagen fibrils under compression.  Studies of collagen fibrils in compressed bone by \citet{Groetsch:2019} have determined yield strains of $\approx 1\%$ with a yield stress of $\sigma/E \approx 0.01$ \cite{bone}. Since from Eqn.~\ref{youngs} the Young's modulus for an elastomer is $E \simeq 3 \mu$ for small $\psi_0$, this gives $\sigma \approx 0.03 \mu$  -- remarkably similar to our estimate.

Polarization-resolved second harmonic generation (P-SHG) microscopy, a technique which takes advantage of non-linear optical phenomena to measure volume-averaged anisotropy within a sample, could potentially be used to measure the change in molecular twist within different types of collagen fibrils under extension or compression. P-SHG anisotropy measurements of collagen-rich tissues such as full tendons have been made in recent years \cite{Gusachenko:2012, Rouede:2020}, however no experimental measurements have yet been made at the single-fibril level. Measurements of twist within strained fibrils would allow our model to be tested further.

Mechanical effects of coexistence may be easier to detect from detailed stress-strain curves. At coexistence, substantial compressive strains can be realized at constant stress -- as detailed in Fig.~\ref{fig:coexistence} and \ref{fig:pvstate}. Short biaxially strained cylinders embedded within an elastic substrate \cite{Schmidtke:2005} provide an accessible geometry for exploring the compression effects we describe here. Long and unsupported double-twist cylinders under compression would exhibit Euler buckling, which might complicate axial strain application and estimation.  However, lateral reinforcement can significantly increase the load that can be applied to long cylinders without buckling \cite{Brangwynne:2006}. As discussed above, compressive studies have been done by attaching single fibrils to pre-stretched elastic substrates \cite{Peacock:2020} though stress is not easily measured in that configuration. 

Strain coexistence implies that applying compressive strains above $\epsilon_L$ leads to compressive strains at $\epsilon_H$. We call this ``strain-leveraging'' -- after the lever-rule of phase-coexistence. As such, very high strains may be achieved for isolated double-twist cylinders. Since this may also lead to plastic damage \cite{Peacock:2020}, it is interesting to speculate that \emph{in vivo} ultrastructure -- such as closely packed fibrils within tendon -- may suppress chiral buckling much as it suppresses Euler buckling \cite{Brangwynne:2006} -- due to the radial expansion exhibited by these instabilities. 

While our focus is on thermodynamic equilibrium elastomeric configurations, an interesting dynamical consequence would be observed if the system was rapidly mechanically ``quenched'' to a compressive strain within the spinodal region, with $\partial^2 f/\partial \lambda^2<0$, where the system will spontaneously nucleate both phases. We would expect interfaces between the coexisting phases to have excess free-energy, as described by a more detailed treatment that includes gradient terms in the director field and characterized by Frank elastic coefficients \cite{Leighton:2021a}.  To reduce such interfacial costs, the system would then slowly ``coarsen'' towards bulk coexistence. Such coarsening could be exceptionally slow since it would not be driven by interfacial curvature \cite{Bray:2002} but instead by exponentially-small interactions between distant interfaces \cite{Rutenberg:1994}. As such it may be experimentally accessible, and could provide details of the non-elastomeric contributions to the free-energy.

Two open questions remain in terms of our calculation. The first is the effect of the Frank free-energy terms for a weakly cross-linked elastomer. We anticipate a rich phase-diagram under compression. The second is the role of shear deformations when strained fibrils are allowed to freely rotate – as might be expected in some experimental setups. We will address this second question in future work \cite{Leighton:2021b}.

\section*{Author Contributions}
MPL wrote the original draft, reviewed and edited, did the visualization, developed the software, and did formal analysis and numerics. LK reviewed and edited and conceptualized the approach. ADR reviewed and edited, supervised, acquired funding, conceptualized the approach, and contributed some of the formal analysis.

\section*{Conflicts of interest}
There are no conflicts to declare.

\section*{Acknowledgements}
We thank the Natural Sciences and Engineering Research Council of Canada (NSERC) for operating Grants RGPIN-2018-03781 (LK) and RGPIN-2019-05888 (ADR). MPL thanks NSERC for summer fellowship support (USRA-552365-2020), and a CGS Masters fellowship.
\balance

\bibliography{main} 

\providecommand*{\mcitethebibliography}{\thebibliography}
\csname @ifundefined\endcsname{endmcitethebibliography}
{\let\endmcitethebibliography\endthebibliography}{}
\begin{mcitethebibliography}{42}
\providecommand*{\natexlab}[1]{#1}
\providecommand*{\mciteSetBstSublistMode}[1]{}
\providecommand*{\mciteSetBstMaxWidthForm}[2]{}
\providecommand*{\mciteBstWouldAddEndPuncttrue}
  {\def\EndOfBibitem{\unskip.}}
\providecommand*{\mciteBstWouldAddEndPunctfalse}
  {\let\EndOfBibitem\relax}
\providecommand*{\mciteSetBstMidEndSepPunct}[3]{}
\providecommand*{\mciteSetBstSublistLabelBeginEnd}[3]{}
\providecommand*{\EndOfBibitem}{}
\mciteSetBstSublistMode{f}
\mciteSetBstMaxWidthForm{subitem}
{(\emph{\alph{mcitesubitemcount}})}
\mciteSetBstSublistLabelBeginEnd{\mcitemaxwidthsubitemform\space}
{\relax}{\relax}

\bibitem[Bryson \emph{et~al.}(2009)Bryson, Harland, Caldwell, Vernon, Walls,
  Woods, Nagase, Itou, and Koike]{Bryson:2009}
W.~G. Bryson, D.~P. Harland, J.~P. Caldwell, J.~A. Vernon, R.~J. Walls, J.~L.
  Woods, S.~Nagase, T.~Itou and K.~Koike, \emph{Journal of Structural Biology},
  2009, \textbf{166}, 46--58\relax
\mciteBstWouldAddEndPuncttrue
\mciteSetBstMidEndSepPunct{\mcitedefaultmidpunct}
{\mcitedefaultendpunct}{\mcitedefaultseppunct}\relax
\EndOfBibitem
\bibitem[Harland \emph{et~al.}(2011)Harland, Caldwell, Woods, Walls, and
  Bryson]{Harland:2011}
D.~P. Harland, J.~P. Caldwell, J.~L. Woods, R.~J. Walls and W.~G. Bryson,
  \emph{Journal of Structural Biology}, 2011, \textbf{173}, 29--37\relax
\mciteBstWouldAddEndPuncttrue
\mciteSetBstMidEndSepPunct{\mcitedefaultmidpunct}
{\mcitedefaultendpunct}{\mcitedefaultseppunct}\relax
\EndOfBibitem
\bibitem[Sherman \emph{et~al.}(2015)Sherman, Yang, and Meyers]{Sherman:2015}
V.~R. Sherman, W.~Yang and M.~A. Meyers, \emph{Journal of the Mechanical
  Behavior of Biomedical Materials}, 2015, \textbf{52}, 22--50\relax
\mciteBstWouldAddEndPuncttrue
\mciteSetBstMidEndSepPunct{\mcitedefaultmidpunct}
{\mcitedefaultendpunct}{\mcitedefaultseppunct}\relax
\EndOfBibitem
\bibitem[Raspanti \emph{et~al.}(2018)Raspanti, Reguzzoni, Protasoni, and
  Basso]{Raspanti:2018}
M.~Raspanti, M.~Reguzzoni, M.~Protasoni and P.~Basso, \emph{International
  Journal of Biological Macromolecules}, 2018, \textbf{107}, 1668--1674\relax
\mciteBstWouldAddEndPuncttrue
\mciteSetBstMidEndSepPunct{\mcitedefaultmidpunct}
{\mcitedefaultendpunct}{\mcitedefaultseppunct}\relax
\EndOfBibitem
\bibitem[Wright and Mermin(1989)]{Wright:1989}
D.~C. Wright and N.~D. Mermin, \emph{Reviews of Modern Physics}, 1989,
  \textbf{61}, 385\relax
\mciteBstWouldAddEndPuncttrue
\mciteSetBstMidEndSepPunct{\mcitedefaultmidpunct}
{\mcitedefaultendpunct}{\mcitedefaultseppunct}\relax
\EndOfBibitem
\bibitem[Eekhoff \emph{et~al.}(2018)Eekhoff, Fang, and Lake]{Eekhoff:2018}
J.~D. Eekhoff, F.~Fang and S.~P. Lake, \emph{Connective Tissue Research}, 2018,
  \textbf{59}, 410--422\relax
\mciteBstWouldAddEndPuncttrue
\mciteSetBstMidEndSepPunct{\mcitedefaultmidpunct}
{\mcitedefaultendpunct}{\mcitedefaultseppunct}\relax
\EndOfBibitem
\bibitem[Leighton \emph{et~al.}(2021)Leighton, Kreplak, and
  Rutenberg]{Leighton:2021a}
M.~P. Leighton, L.~Kreplak and A.~D. Rutenberg, \emph{Soft Matter}, 2021,
  https://doi.org/10.1039/D0SM01830A\relax
\mciteBstWouldAddEndPuncttrue
\mciteSetBstMidEndSepPunct{\mcitedefaultmidpunct}
{\mcitedefaultendpunct}{\mcitedefaultseppunct}\relax
\EndOfBibitem
\bibitem[Matoltsy(1976)]{Matoltsy:1976}
A.~G. Matoltsy, \emph{The Journal of Investigative Dermatology}, 1976,
  \textbf{67}, 20--25\relax
\mciteBstWouldAddEndPuncttrue
\mciteSetBstMidEndSepPunct{\mcitedefaultmidpunct}
{\mcitedefaultendpunct}{\mcitedefaultseppunct}\relax
\EndOfBibitem
\bibitem[Gautieri \emph{et~al.}(2017)Gautieri, Passini, Silv{\'a}n,
  Guizar-Sicairos, Carimati, Volpi, Moretti, Schoenhuber, Redaelli, Berli, and
  Snedeker]{Gautieri:2017}
A.~Gautieri, F.~S. Passini, U.~Silv{\'a}n, M.~Guizar-Sicairos, G.~Carimati,
  P.~Volpi, M.~Moretti, H.~Schoenhuber, A.~Redaelli, M.~Berli and J.~G.
  Snedeker, \emph{Matrix Biology}, 2017, \textbf{59}, 95--108\relax
\mciteBstWouldAddEndPuncttrue
\mciteSetBstMidEndSepPunct{\mcitedefaultmidpunct}
{\mcitedefaultendpunct}{\mcitedefaultseppunct}\relax
\EndOfBibitem
\bibitem[Warner and Terentjev(1996)]{Warner:1996}
M.~Warner and E.~Terentjev, \emph{Progress in Polymer Science}, 1996,
  \textbf{21}, 853--891\relax
\mciteBstWouldAddEndPuncttrue
\mciteSetBstMidEndSepPunct{\mcitedefaultmidpunct}
{\mcitedefaultendpunct}{\mcitedefaultseppunct}\relax
\EndOfBibitem
\bibitem[Warner \emph{et~al.}(2000)Warner, Terentjev, Meyer, and
  Mao]{Warner:2000}
M.~Warner, E.~Terentjev, R.~Meyer and Y.~Mao, \emph{Physical Review Letters},
  2000, \textbf{85}, 2320\relax
\mciteBstWouldAddEndPuncttrue
\mciteSetBstMidEndSepPunct{\mcitedefaultmidpunct}
{\mcitedefaultendpunct}{\mcitedefaultseppunct}\relax
\EndOfBibitem
\bibitem[Menzel and Brand(2007)]{Menzel:2007}
A.~M. Menzel and H.~R. Brand, \emph{Physical Review E}, 2007, \textbf{75},
  011707\relax
\mciteBstWouldAddEndPuncttrue
\mciteSetBstMidEndSepPunct{\mcitedefaultmidpunct}
{\mcitedefaultendpunct}{\mcitedefaultseppunct}\relax
\EndOfBibitem
\bibitem[Stille(2009)]{Stille:2009}
W.~Stille, \emph{The European Physical Journal E}, 2009, \textbf{28},
  57--71\relax
\mciteBstWouldAddEndPuncttrue
\mciteSetBstMidEndSepPunct{\mcitedefaultmidpunct}
{\mcitedefaultendpunct}{\mcitedefaultseppunct}\relax
\EndOfBibitem
\bibitem[Xing and Baskaran(2008)]{Xing:2008}
X.~Xing and A.~Baskaran, \emph{Physical Review E}, 2008, \textbf{78},
  021709\relax
\mciteBstWouldAddEndPuncttrue
\mciteSetBstMidEndSepPunct{\mcitedefaultmidpunct}
{\mcitedefaultendpunct}{\mcitedefaultseppunct}\relax
\EndOfBibitem
\bibitem[Finkelmann(2001)]{Finkelmann:2001}
H.~Finkelmann, \emph{Advanced Materials}, 2001, \textbf{13}, 1069--1072\relax
\mciteBstWouldAddEndPuncttrue
\mciteSetBstMidEndSepPunct{\mcitedefaultmidpunct}
{\mcitedefaultendpunct}{\mcitedefaultseppunct}\relax
\EndOfBibitem
\bibitem[Biscari \emph{et~al.}(2014)Biscari, DiCarlo, and Turzi]{Biscari:2014}
P.~Biscari, A.~DiCarlo and S.~S. Turzi, \emph{Soft Matter}, 2014, \textbf{10},
  8296--8307\relax
\mciteBstWouldAddEndPuncttrue
\mciteSetBstMidEndSepPunct{\mcitedefaultmidpunct}
{\mcitedefaultendpunct}{\mcitedefaultseppunct}\relax
\EndOfBibitem
\bibitem[Singamaneni and Tsukruk(2010)]{Singamaneni:2010}
S.~Singamaneni and V.~V. Tsukruk, \emph{Soft Matter}, 2010, \textbf{6},
  5681--5692\relax
\mciteBstWouldAddEndPuncttrue
\mciteSetBstMidEndSepPunct{\mcitedefaultmidpunct}
{\mcitedefaultendpunct}{\mcitedefaultseppunct}\relax
\EndOfBibitem
\bibitem[Rey \emph{et~al.}(2016)Rey, Fern{\'a}ndez-Rodr{\'\i}guez, Steinacher,
  Scheidegger, Geisel, Richtering, Squires, and Isa]{Rey:2016}
M.~Rey, M.~{\'A}. Fern{\'a}ndez-Rodr{\'\i}guez, M.~Steinacher, L.~Scheidegger,
  K.~Geisel, W.~Richtering, T.~M. Squires and L.~Isa, \emph{Soft Matter}, 2016,
  \textbf{12}, 3545--3557\relax
\mciteBstWouldAddEndPuncttrue
\mciteSetBstMidEndSepPunct{\mcitedefaultmidpunct}
{\mcitedefaultendpunct}{\mcitedefaultseppunct}\relax
\EndOfBibitem
\bibitem[Rafsanjani \emph{et~al.}(2019)Rafsanjani, Jin, Deng, and
  Bertoldi]{Rafsanjani:2019}
A.~Rafsanjani, L.~Jin, B.~Deng and K.~Bertoldi, \emph{Proceedings of the
  National Academy of Sciences}, 2019, \textbf{116}, 8200--8205\relax
\mciteBstWouldAddEndPuncttrue
\mciteSetBstMidEndSepPunct{\mcitedefaultmidpunct}
{\mcitedefaultendpunct}{\mcitedefaultseppunct}\relax
\EndOfBibitem
\bibitem[Miyazaki and Kondo(1997)]{Miyazaki:1997}
Y.~Miyazaki and K.~Kondo, \emph{International Journal of Solids and
  Structures}, 1997, \textbf{34}, 3619--3636\relax
\mciteBstWouldAddEndPuncttrue
\mciteSetBstMidEndSepPunct{\mcitedefaultmidpunct}
{\mcitedefaultendpunct}{\mcitedefaultseppunct}\relax
\EndOfBibitem
\bibitem[Peacock \emph{et~al.}(2020)Peacock, Lee, Beral, Cisek, Tokarz, and
  Kreplak]{Peacock:2020}
C.~Peacock, E.~Lee, T.~Beral, R.~Cisek, D.~Tokarz and L.~Kreplak, \emph{ACS
  Nano}, 2020, \textbf{14}, 12877--12884\relax
\mciteBstWouldAddEndPuncttrue
\mciteSetBstMidEndSepPunct{\mcitedefaultmidpunct}
{\mcitedefaultendpunct}{\mcitedefaultseppunct}\relax
\EndOfBibitem
\bibitem[Ottani \emph{et~al.}(2001)Ottani, Raspanti, and Ruggeri]{Ottani:2001}
V.~Ottani, M.~Raspanti and A.~Ruggeri, \emph{Micron}, 2001, \textbf{32},
  251--260\relax
\mciteBstWouldAddEndPuncttrue
\mciteSetBstMidEndSepPunct{\mcitedefaultmidpunct}
{\mcitedefaultendpunct}{\mcitedefaultseppunct}\relax
\EndOfBibitem
\bibitem[Holmes \emph{et~al.}(2001)Holmes, Gilpin, Baldock, Ziese, Koster, and
  Kadler]{Holmes:2001}
D.~F. Holmes, C.~J. Gilpin, C.~Baldock, U.~Ziese, A.~J. Koster and K.~E.
  Kadler, \emph{Proceedings of the National Academy of Sciences}, 2001,
  \textbf{98}, 7307--7312\relax
\mciteBstWouldAddEndPuncttrue
\mciteSetBstMidEndSepPunct{\mcitedefaultmidpunct}
{\mcitedefaultendpunct}{\mcitedefaultseppunct}\relax
\EndOfBibitem
\bibitem[Cameron \emph{et~al.}(2020)Cameron, Kreplak, and
  Rutenberg]{Cameron:2020}
S.~Cameron, L.~Kreplak and A.~D. Rutenberg, \emph{Physical Review Research},
  2020, \textbf{2}, 012070\relax
\mciteBstWouldAddEndPuncttrue
\mciteSetBstMidEndSepPunct{\mcitedefaultmidpunct}
{\mcitedefaultendpunct}{\mcitedefaultseppunct}\relax
\EndOfBibitem
\bibitem[d'Allest \emph{et~al.}(1988)d'Allest, Maissa, Ten~Bosch, Sixou,
  Blumstein, Blumstein, Teixeira, and Noirez]{DAllest:1988}
J.~d'Allest, P.~Maissa, A.~Ten~Bosch, P.~Sixou, A.~Blumstein, R.~Blumstein,
  J.~Teixeira and L.~Noirez, \emph{Physical Review Letters}, 1988, \textbf{61},
  2562\relax
\mciteBstWouldAddEndPuncttrue
\mciteSetBstMidEndSepPunct{\mcitedefaultmidpunct}
{\mcitedefaultendpunct}{\mcitedefaultseppunct}\relax
\EndOfBibitem
\bibitem[Kundler and Finkelmann(1998)]{Kundler:1998}
I.~Kundler and H.~Finkelmann, \emph{Macromolecular Chemistry and Physics},
  1998, \textbf{199}, 677--686\relax
\mciteBstWouldAddEndPuncttrue
\mciteSetBstMidEndSepPunct{\mcitedefaultmidpunct}
{\mcitedefaultendpunct}{\mcitedefaultseppunct}\relax
\EndOfBibitem
\bibitem[Landau and Lifshitz(1986)]{Landau:1960}
L.~Landau and E.~Lifshitz, \emph{Theory of Elasticity}, Butterworth-Heinemann,
  3rd edn, 1986\relax
\mciteBstWouldAddEndPuncttrue
\mciteSetBstMidEndSepPunct{\mcitedefaultmidpunct}
{\mcitedefaultendpunct}{\mcitedefaultseppunct}\relax
\EndOfBibitem
\bibitem[git()]{github}
\emph{Github code},
  \url{https://github.com/Matthew-Leighton/Elastomer_Phase_Transitions}\relax
\mciteBstWouldAddEndPuncttrue
\mciteSetBstMidEndSepPunct{\mcitedefaultmidpunct}
{\mcitedefaultendpunct}{\mcitedefaultseppunct}\relax
\EndOfBibitem
\bibitem[mul()]{multicritical}
The coexistence of $\epsilon_H$ and $\epsilon_L$ represents discontinuous phase
  transitions at a given $\zeta$. With respect to $\zeta$, coexistence ends
  above $\zeta_{max}(\psi_0)$ and below $\zeta_{min}(\psi_0) \geq1$. There is a
  mean-field multicritical point at $\psi_0^c \simeq 0.42385$, where
  $\zeta_{max}=\zeta_{min}\simeq 1.002$.\relax
\mciteBstWouldAddEndPunctfalse
\mciteSetBstMidEndSepPunct{\mcitedefaultmidpunct}
{}{\mcitedefaultseppunct}\relax
\EndOfBibitem
\bibitem[Lei()]{Leighton:2021b}
M Leighton, A Rutenberg, and L Kreplak (unpublished)\relax
\mciteBstWouldAddEndPuncttrue
\mciteSetBstMidEndSepPunct{\mcitedefaultmidpunct}
{\mcitedefaultendpunct}{\mcitedefaultseppunct}\relax
\EndOfBibitem
\bibitem[Mitchell \emph{et~al.}(1993)Mitchell, Davis, and Guo]{Mitchell:1993}
G.~R. Mitchell, F.~J. Davis and W.~Guo, \emph{Physical Review Letters}, 1993,
  \textbf{71}, 2947--2950\relax
\mciteBstWouldAddEndPuncttrue
\mciteSetBstMidEndSepPunct{\mcitedefaultmidpunct}
{\mcitedefaultendpunct}{\mcitedefaultseppunct}\relax
\EndOfBibitem
\bibitem[Burridge \emph{et~al.}(2006)Burridge, Mao, and Warner]{Burridge:2006}
D.~J. Burridge, Y.~Mao and M.~Warner, \emph{Physical Review E}, 2006,
  \textbf{74}, 2515\relax
\mciteBstWouldAddEndPuncttrue
\mciteSetBstMidEndSepPunct{\mcitedefaultmidpunct}
{\mcitedefaultendpunct}{\mcitedefaultseppunct}\relax
\EndOfBibitem
\bibitem[Bell \emph{et~al.}(2018)Bell, Hayes, Whitford, Sanchez-Weatherby,
  Shebanova, Vergari, Winlove, Terrill, Sorensen,
  Elsheikh,\emph{et~al.}]{Bell:2018}
J.~Bell, S.~Hayes, C.~Whitford, J.~Sanchez-Weatherby, O.~Shebanova, C.~Vergari,
  C.~Winlove, N.~Terrill, T.~Sorensen, A.~Elsheikh \emph{et~al.}, \emph{Acta
  Biomaterialia}, 2018, \textbf{65}, 216--225\relax
\mciteBstWouldAddEndPuncttrue
\mciteSetBstMidEndSepPunct{\mcitedefaultmidpunct}
{\mcitedefaultendpunct}{\mcitedefaultseppunct}\relax
\EndOfBibitem
\bibitem[Lillie \emph{et~al.}(1977)Lillie, MacCallum, Scaletta, and
  Occhino]{Lillie:1977}
J.~H. Lillie, D.~K. MacCallum, L.~J. Scaletta and J.~C. Occhino, \emph{Journal
  of Ultrasructure Research}, 1977, \textbf{58}, 134--143\relax
\mciteBstWouldAddEndPuncttrue
\mciteSetBstMidEndSepPunct{\mcitedefaultmidpunct}
{\mcitedefaultendpunct}{\mcitedefaultseppunct}\relax
\EndOfBibitem
\bibitem[Groetsch \emph{et~al.}(2019)Groetsch, Gourrier, Schwiedrzik, Sztucki,
  Beck, Shephard, Michler, Zysset, and Wolfram]{Groetsch:2019}
A.~Groetsch, A.~Gourrier, J.~Schwiedrzik, M.~Sztucki, R.~J. Beck, J.~D.
  Shephard, J.~Michler, P.~K. Zysset and U.~Wolfram, \emph{Acta Biomaterialia},
  2019, \textbf{89}, 313--329\relax
\mciteBstWouldAddEndPuncttrue
\mciteSetBstMidEndSepPunct{\mcitedefaultmidpunct}
{\mcitedefaultendpunct}{\mcitedefaultseppunct}\relax
\EndOfBibitem
\bibitem[bon()]{bone}
In bone, \citet{Groetsch:2019} reports $\epsilon^{yield}_{fiber} \approx 4-6\%$
  for the fiber, while $\epsilon^{yield}_{fibril} \approx 0.2
  \epsilon^{yield}_{fiber}$. They also report $\sigma^{yield}_{fiber}/E^{fiber}
  \approx 0.01$.\relax
\mciteBstWouldAddEndPunctfalse
\mciteSetBstMidEndSepPunct{\mcitedefaultmidpunct}
{}{\mcitedefaultseppunct}\relax
\EndOfBibitem
\bibitem[Gusachenko \emph{et~al.}(2012)Gusachenko, Tran, Houssen, Allain, and
  Schanne-Klein]{Gusachenko:2012}
I.~Gusachenko, V.~Tran, Y.~G. Houssen, J.-M. Allain and M.-C. Schanne-Klein,
  \emph{Biophysical Journal}, 2012, \textbf{102}, 2220--2229\relax
\mciteBstWouldAddEndPuncttrue
\mciteSetBstMidEndSepPunct{\mcitedefaultmidpunct}
{\mcitedefaultendpunct}{\mcitedefaultseppunct}\relax
\EndOfBibitem
\bibitem[Rou{\`e}de \emph{et~al.}(2020)Rou{\`e}de, Schaub, Bellanger, Ezan, and
  Tiaho]{Rouede:2020}
D.~Rou{\`e}de, E.~Schaub, J.-J. Bellanger, F.~Ezan and F.~Tiaho, \emph{Optics
  Express}, 2020, \textbf{28}, 4845--4858\relax
\mciteBstWouldAddEndPuncttrue
\mciteSetBstMidEndSepPunct{\mcitedefaultmidpunct}
{\mcitedefaultendpunct}{\mcitedefaultseppunct}\relax
\EndOfBibitem
\bibitem[Schmidtke \emph{et~al.}(2005)Schmidtke, Kniesel, and
  Finkelmann]{Schmidtke:2005}
J.~Schmidtke, S.~Kniesel and H.~Finkelmann, \emph{Macromolecules}, 2005,
  \textbf{38}, 1357--1363\relax
\mciteBstWouldAddEndPuncttrue
\mciteSetBstMidEndSepPunct{\mcitedefaultmidpunct}
{\mcitedefaultendpunct}{\mcitedefaultseppunct}\relax
\EndOfBibitem
\bibitem[Brangwynne \emph{et~al.}(2006)Brangwynne, Mackintosh, Kumar, Geisse,
  Talbot, Mahadevan, Parker, Ingber, and Weitz]{Brangwynne:2006}
C.~P. Brangwynne, F.~C. Mackintosh, S.~Kumar, N.~Geisse, J.~Talbot,
  L.~Mahadevan, K.~Parker, D.~Ingber and D.~A. Weitz, \emph{The Journal of Cell
  Biology}, 2006, \textbf{173}, 733--741\relax
\mciteBstWouldAddEndPuncttrue
\mciteSetBstMidEndSepPunct{\mcitedefaultmidpunct}
{\mcitedefaultendpunct}{\mcitedefaultseppunct}\relax
\EndOfBibitem
\bibitem[Bray(2002)]{Bray:2002}
A.~J. Bray, \emph{Advances in Physics}, 2002, \textbf{51}, 481--587\relax
\mciteBstWouldAddEndPuncttrue
\mciteSetBstMidEndSepPunct{\mcitedefaultmidpunct}
{\mcitedefaultendpunct}{\mcitedefaultseppunct}\relax
\EndOfBibitem
\bibitem[Rutenberg and Bray(1994)]{Rutenberg:1994}
A.~D. Rutenberg and A.~J. Bray, \emph{Physical Review E}, 1994, \textbf{50},
  1900--1911\relax
\mciteBstWouldAddEndPuncttrue
\mciteSetBstMidEndSepPunct{\mcitedefaultmidpunct}
{\mcitedefaultendpunct}{\mcitedefaultseppunct}\relax
\EndOfBibitem
\end{mcitethebibliography}
\bibliographystyle{rsc} 

\end{document}